\documentclass[11pt]{article}
\usepackage[margin=0.88in]{geometry}
\usepackage{amsmath,amssymb,amsthm,mathtools,stmaryrd}
\usepackage{microtype}
\usepackage{booktabs,array,tabularx}
\usepackage{enumitem}
\usepackage{tikz}
\usetikzlibrary{arrows.meta,positioning,calc,fit,backgrounds,matrix,decorations.pathreplacing}
\usepackage{xcolor}
\definecolor{linkblue}{RGB}{24,76,130}
\definecolor{theoremblue}{RGB}{238,246,252}
\definecolor{figblue}{RGB}{45,96,150}
\definecolor{figgreen}{RGB}{54,117,94}
\definecolor{figred}{RGB}{158,63,66}
\definecolor{figgold}{RGB}{156,117,42}
\definecolor{figgray}{RGB}{92,101,111}
\definecolor{softgray}{RGB}{246,247,249}
\usepackage[colorlinks=true,linkcolor=linkblue,citecolor=linkblue,urlcolor=linkblue]{hyperref}
\usepackage[nameinlink,capitalize]{cleveref}
\usepackage{mathpazo}
\usepackage[T1]{fontenc}
\usepackage{authblk}
\usepackage{caption}
\setlist{nosep,leftmargin=*}
\newtheorem{theorem}{Theorem}[section]

\newtheorem{proposition}[theorem]{Proposition}
\newtheorem{corollary}[theorem]{Corollary}
\newtheorem{definition}[theorem]{Definition}

\newcommand{\R}{\mathbb R}
\newcommand{\cH}{\mathcal H}

\newcommand{\cI}{\mathcal I}
\newcommand{\cU}{\mathcal U}
\newcommand{\Fix}{\operatorname{Fix}}
\newcommand{\Norm}{\operatorname{N}}
\newcommand{\Prog}{\operatorname{Prog}}

\newcommand{\Sym}{\operatorname{Sym}}
\newcommand{\ket}[1]{\lvert #1\rangle}

\newcommand{\proj}[1]{\lvert #1\rangle\!\langle #1\rvert}

\title{\bfseries Fixed Points as Programs: Finite Symmetry and One-Shot State–Gate Duality in the Misty-State Calculus}
\author[1]{Dan-Adrian German}
\author[2]{Tony Newton\thanks{Newton Astro Labs is the trading name under which Tony Newton conducts independent computational research in the United Kingdom; it is not a limited company.}}
\affil[1]{Luddy SICE, Indiana University, Bloomington, IN, USA\\\texttt{dgerman@iu.edu}}
\affil[2]{Newton Astro Labs*, London, UK\\\texttt{tony.newton79@gmail.com}}
\date{October 2026}

\begin{document}
\maketitle

\begin{abstract}
The same quantum object can support several logically distinct descriptions: it can be an eigenstate of an operator, a fixed point of an action, a nonstabilizer resource, or a state consumed by a gate-injection protocol.  We make these relations precise in the misty-state calculus.  The semantic core is a normalized superposition constructor $x\star y=\Norm(x+y)$.  For any finite unitary group $G$, whenever the orbit sum is nonzero, the normalized orbit cloud $\Norm(\sum_{g\in G}U_gx)$ is fixed by every element of $G$; the involutive specialization $x\star Ux$ is therefore a certified fixed-point constructor.  Applied to $U=H$ and $x=\ket0$, it gives the nested misty state
\[
 M=\Norm\{W,\Norm\{W,B\}\}=\cos(\pi/8)\ket0+\sin(\pi/8)\ket1,
\]
the $+1$ Hadamard eigenstate.  This state is not a Pauli stabilizer state; it is an H-type magic state and is Clifford-equivalent to $T\ket+$.  We then prove an exact one-shot state--gate duality.  Writing $J=XZ$ and $R_\theta=e^{\theta J}$, the state $M_\theta=R_\theta\ket0$ can be consumed by a Clifford controlled-$J$ injection circuit to apply either $R_\theta$ or $R_\theta^{-1}$ to arbitrary data.  Whenever $R_\theta^2$ is free, feed-forward makes the implementation deterministic.  At $\theta=\pi/8$, $R_\theta^2=HZ=XH$ is Clifford, so one copy of $M$ implements one exact use of the non-Clifford gate $R=e^{\pi J/8}$; conversely one use of $R$ prepares $M$ from $\ket0$.  We formulate this as a pair of sound semantic certificates, $\Fix(U)$ and $\Prog(R)$, and separate the resulting finite symmetrization mechanism from the genuinely recursive fixed-point role of the $Y$ combinator, while identifying the exact $S$-combinator sharing pattern in $x\star Ux$.  The result is a small, auditable semantics in which invariance and consumable program resources coexist without being conflated.
\end{abstract}

\section{Introduction and main results}\label{sec:intro}

A recurring pattern across quantum information and programming-language semantics is that an operation acts on an object while some structure survives.  The equations look deceptively similar.  An eigenvector satisfies $Av=\lambda v$; a projective quantum state is fixed when $[U\psi]=[\psi]$; a stabilizer state is fixed by a family of Pauli operators; a fixed-point combinator constructs a solution of $f(x)=x$; an attractor is a fixed object approached dynamically.  These notions are related, but identifying them would erase exactly the distinctions that make them useful.

The misty-state language gives a compact setting in which the distinctions can be made explicit.  Rudolph's original black-and-white formalism represents basis states by tokens and superpositions by clouds, with interference realized by symbolic cancellation \cite{Rudolph2017}.  German recast the language as a term-rewriting semantics and introduced irreducible nested states with fixed-point behavior \cite{German2026}.  Newton and German subsequently separated raw amplitude addition from physical normalization and supplied an exact, falsifiable semantics for finite mist expressions \cite{NewtonGerman2026}.  The present paper develops the fixed-point and resource-program consequences of that semantic separation.

The starting object is the nested state
\begin{equation}\label{eq:Mintro}
  M=\Norm\{W,\Norm\{W,B\}\}.
\end{equation}
Its significance is twofold.  First, it is an exact fixed point of Hadamard.  Second, it is a nonstabilizer state that can be consumed to apply a non-Clifford rotation to arbitrary data.  The phrase ``state as program'' is therefore not used metaphorically here: we give a concrete processor architecture and prove its induced channel.

\paragraph{Main results.}
The paper establishes four claims.
\begin{enumerate}[label=(\roman*)]
\item A finite orbit cloud is an exact fixed-point constructor.  The involution case $x\star Ux$ gives a particularly small certified constructor and places the nested Hadamard state inside a general symmetry principle.
\item The state in \eqref{eq:Mintro} is simultaneously a Hadamard fixed point and an H-type magic state, while not being a Pauli stabilizer state.  This provides a precise example of why eigenstate, stabilizer state, and magic state must not be conflated.
\item A one-parameter injection identity converts $M_\theta=e^{\theta J}\ket0$ into the gate $R_\theta=e^{\theta J}$ on arbitrary data.  If $R_\theta^2$ is free, the conversion is deterministic.  At $\theta=\pi/8$, the correction is Clifford, giving an exact one-shot equivalence between one copy of $M$ and one invocation of $R$ relative to Clifford processing.
\item The fixed-point constructor has the exact sharing pattern $S\star Ux=\star x(Ux)$ of the $S$ combinator, but it is finite orbit symmetrization rather than recursive self-application.  The $Y$ combinator supplies the genuinely recursive equation $Yf=f(Yf)$.  We encode this distinction by separate semantic certificates for invariance and resource-program behavior.
\end{enumerate}

\begin{figure}[t]
\centering
\begin{tikzpicture}[node distance=8mm and 12mm, every node/.style={font=\small},
 box/.style={draw=black,rounded corners=2pt,fill=white,minimum height=9mm,align=center,inner sep=4pt},
 arr/.style={-{Latex[length=2.4mm]},thick,draw=figblue}]
\node[box,fill=theoremblue] (m) {$M=\Norm\{W,\Norm\{W,B\}\}$};
\node[box,above right=of m, yshift=-5mm] (inv) {invariance certificate\\$HM=M$};
\node[box,below right=of m, yshift=5mm] (prog) {program certificate\\$M\leadsto R$ on data};
\node[box,right=of inv] (fix) {$M:\Fix(H)$};
\node[box,right=of prog] (r) {$M:\Prog(R)$};
\draw[arr] (m)--(inv); \draw[arr] (m)--(prog);
\draw[arr] (inv)--(fix); \draw[arr] (prog)--(r);
\end{tikzpicture}
\caption{One state supports two different certificates.  The upper route is an equation in state semantics; the lower route is an operational statement about a processor that consumes the state.}
\label{fig:twofaces}
\end{figure}
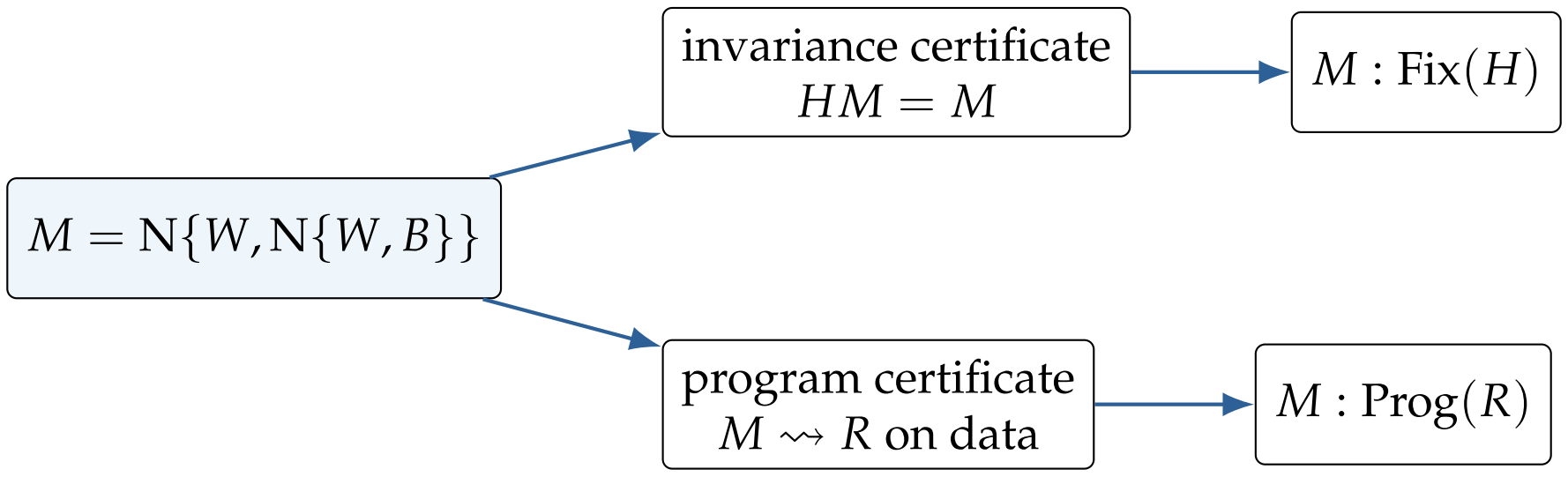

The underlying ingredients---group averaging, Clifford magic states, and gate injection---are established ideas in their respective subjects \cite{BravyiKitaev2005,GottesmanChuang1999}.  The contribution here is their exact organization inside the misty-state syntax: normalized nesting becomes a certified invariance constructor, while the same denotation becomes a first-class consumable program resource.  We state claim boundaries explicitly in \cref{sec:scope}.

\section{Semantic core of the misty-state calculus}\label{sec:semantics}

We use only the finite pure-state fragment needed for the results.  Let $W$ and $B$ denote the computational basis tokens,
\[
   \llbracket W\rrbracket=\ket0,\qquad
   \llbracket B\rrbracket=\ket1.
\]
For signed finite clouds,
\begin{align*}
 \llbracket -t\rrbracket&=-\llbracket t\rrbracket,\\
 \llbracket\{t_1,\ldots,t_m\}\rrbracket&=\sum_{j=1}^m\llbracket t_j\rrbracket.
\end{align*}
Normalization is an explicit semantic operation
\begin{equation}\label{eq:norm}
 \llbracket \Norm(t)\rrbracket
 =\frac{\llbracket t\rrbracket}{\|\llbracket t\rrbracket\|},
 \qquad \llbracket t\rrbracket\ne0.
\end{equation}
This scope discipline matters.  For example,
\[
 \Norm\{W,\{W,B\}\}
 \quad\text{and}\quad
 \Norm\{W,\Norm\{W,B\}\}
\]
are not equivalent: the first normalizes a raw amplitude sum proportional to $2\ket0+\ket1$, whereas the second first converts the inner cloud into a unit state.

\subsection{Normalized combination}
For nonzero vectors $x+y$, define
\begin{equation}\label{eq:star}
  x\star y:=\Norm(x+y).
\end{equation}
The operation is symmetric and equivariant under every unitary $U$:
\begin{equation}\label{eq:equivariance}
 U(x\star y)=Ux\star Uy.
\end{equation}
The equality follows because unitaries preserve norms.  On the real great-circle family
\[
 q(\alpha)=\cos\alpha\ket0+\sin\alpha\ket1,
 \qquad 0\le \alpha\le \frac\pi2,
\]
we have the midpoint law
\begin{equation}\label{eq:midpoint}
 q(\alpha)\star q(\beta)=q\!\left(\frac{\alpha+\beta}{2}\right).
\end{equation}
Indeed, the usual sum-to-product identities give
\[
 q(\alpha)+q(\beta)
 =2\cos\!\left(\frac{\alpha-\beta}{2}\right)
 q\!\left(\frac{\alpha+\beta}{2}\right),
\]
and the prefactor is positive on the declared interval.

\subsection{The Hadamard state}
Since
\[
 \Norm\{W,B\}=\ket+=q(\pi/4),
\]
\cref{eq:midpoint} gives
\begin{equation}\label{eq:Mangle}
 M=\ket0\star\ket+
  =q(\pi/8)
  =\cos\frac\pi8\ket0+\sin\frac\pi8\ket1.
\end{equation}
Direct multiplication shows
\begin{equation}\label{eq:HM}
 HM=M.
\end{equation}
The same fact will reappear below as a consequence of finite symmetrization rather than a special trigonometric identity.

\section{Finite symmetry as a fixed-point constructor}\label{sec:symmetry}

The fixed-point mechanism extends beyond Hadamard and beyond involutions.  Let a finite group $G$ act unitarily on a Hilbert space through $g\mapsto U_g$.

\begin{definition}[Orbit cloud]\label{def:orbitcloud}
For $x\in\cH$ with nonzero orbit sum, define
\begin{equation}\label{eq:orbitcloud}
 \Sym_G(x)
 :=\Norm\!\left(\sum_{g\in G}U_gx\right).
\end{equation}
\end{definition}

\begin{theorem}[Finite-orbit invariance]\label{thm:orbit}
For every $h\in G$,
\[
 U_h\Sym_G(x)=\Sym_G(x).
\]
Thus the orbit cloud is a certified common fixed point of the group action whenever the orbit sum is nonzero.
\end{theorem}

\begin{proof}
Left multiplication permutes the elements of $G$:
\[
 U_h\sum_{g\in G}U_gx
 =\sum_{g\in G}U_{hg}x
 =\sum_{g'\in G}U_{g'}x.
\]
The vector is therefore unchanged before normalization, and hence after normalization.
\end{proof}

For an involution $U^2=I$, the group is $\{I,U\}$ and
\begin{equation}\label{eq:fixstar}
 \Fix_\star(U,x):=x\star Ux
\end{equation}
is a fixed point whenever $x+Ux\ne0$.

\begin{corollary}[Involution constructor]\label{cor:involution}
If $U^2=I$ and $x+Ux\ne0$, then
\[
 U\Fix_\star(U,x)=\Fix_\star(U,x).
\]
\end{corollary}

\begin{proof}
Using \cref{eq:equivariance}, symmetry of $\star$, and $U^2=I$,
\[
 U(x\star Ux)=Ux\star U^2x=Ux\star x=x\star Ux.
\]
\end{proof}

The exceptional condition is essential: if $Ux=-x$, the orbit sum vanishes and normalization is undefined.  In operator language the unnormalized two-element orbit sum is $(I+U)x=2P_+x$, where $P_+=(I+U)/2$ is the projector onto the $+1$ eigenspace.  Thus the constructor is not an attractor process; it is a finite algebraic projection followed by normalization.

Taking $U=H$ and $x=\ket0$ gives
\[
 \Fix_\star(H,\ket0)=\ket0\star H\ket0
 =\ket0\star\ket+=M,
\]
so \cref{eq:HM} is a special case of \cref{cor:involution}.

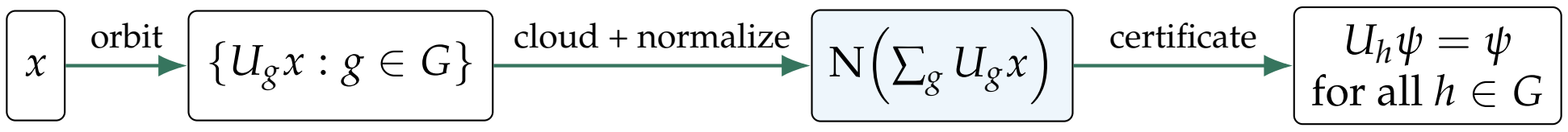
\begin{figure}[t]
\centering
\begin{tikzpicture}[node distance=8mm and 12mm, every node/.style={font=\small},
 box/.style={draw=black,rounded corners=2pt,fill=white,minimum height=9mm,align=center,inner sep=4pt},
 arr/.style={-{Latex[length=2.4mm]},thick,draw=figgreen}]
\node[box] (x) {$x$};
\node[box,right=of x, xshift=-2mm] (orbit) {$\{U_gx:g\in G\}$};
\node[box,right=of orbit,fill=theoremblue, xshift=14mm] (sum) {$\Norm\!\left(\sum_gU_gx\right)$};
\node[box,right=of sum, xshift=6mm] (fixed) {$U_h\psi=\psi$\\for all $h\in G$};
\draw[arr] (x)-- node[above,font=\scriptsize]{orbit} (orbit);
\draw[arr] (orbit)-- node[above,font=\scriptsize]{cloud + normalize} (sum);
\draw[arr] (sum)-- node[above,font=\scriptsize]{certificate} (fixed);
\end{tikzpicture}
\caption{Finite orbit symmetrization as a fixed-point constructor.  The theorem is purely algebraic; no recursive limit or dynamical convergence is involved.}
\label{fig:orbit}
\end{figure}

\section{The Hadamard fixed point as a magic resource}\label{sec:magic}

The state $M$ is a useful test of conceptual boundaries because three classifications intersect.

\subsection{Eigenstate but not Pauli stabilizer state}
Equation \eqref{eq:HM} makes $M$ a $+1$ eigenstate of the Clifford gate $H$.  This does \emph{not} make it a stabilizer state in the usual qubit Pauli-stabilizer sense.  A pure one-qubit Pauli stabilizer state has Bloch vector on one of the coordinate axes.  From \eqref{eq:Mangle},
\[
 \vec r(M)=\left(\sin\frac\pi4,0,\cos\frac\pi4\right)
 =\left(\frac1{\sqrt2},0,\frac1{\sqrt2}\right),
\]
which is not a coordinate-axis vector.  Thus $M$ is a fixed point of a Clifford action while remaining nonstabilizer.

This distinction is central to magic-state computation.  Bravyi and Kitaev identify pure nonstabilizer states in special directions as magic resources which, combined with Clifford processing, enable universal quantum computation \cite{BravyiKitaev2005}.

\subsection{Clifford equivalence to the phase magic state}
Let
\[
 S=\begin{pmatrix}1&0\\0&i\end{pmatrix},
 \qquad
 T=\begin{pmatrix}1&0\\0&e^{i\pi/4}\end{pmatrix},
 \qquad
 \ket+=\frac{\ket0+\ket1}{\sqrt2}.
\]
Then
\begin{proposition}[H-type/phase-type equivalence]\label{prop:cliffordequiv}
The misty Hadamard state is Clifford-equivalent to $T\ket+$:
\begin{equation}\label{eq:cliffordequiv}
 HS^\dagger M=e^{-i\pi/8}T\ket+.
\end{equation}
\end{proposition}

\begin{proof}
Writing $c=\cos(\pi/8)$ and $s=\sin(\pi/8)$,
\[
 HS^\dagger M
 =\frac1{\sqrt2}\begin{pmatrix}c-is\\c+is\end{pmatrix}
 =\frac{e^{-i\pi/8}}{\sqrt2}
   \begin{pmatrix}1\\e^{i\pi/4}\end{pmatrix}
 =e^{-i\pi/8}T\ket+.
\]
The leading phase is physically irrelevant.
\end{proof}

Hence the fixed-point state discovered syntactically in the misty calculus lies on a standard one-qubit magic-state orbit.  The mathematical facts are established; the new semantic point is that the same misty term can carry an invariance certificate and a program-resource certificate simultaneously.

\begin{table}[t]
\centering
\caption{Closely related notions remain logically distinct.}
\label{tab:taxonomy}
\small
\begin{tabularx}{\textwidth}{@{}lXl@{}}
\toprule
Notion & Defining statement & Status of $M$ \\
\midrule
Eigenstate & $A\psi=\lambda\psi$ & Yes: $HM=M$ \\
Projective fixed point & $[U\psi]=[\psi]$ & Yes \\
Pauli stabilizer state & $P\psi=\psi$ for a maximal Pauli subgroup & No \\
Magic state & nonstabilizer resource for free Clifford processing & Yes \\
Attractor & neighboring states converge under dynamics & Not implied by $HM=M$ \\
Program resource & a processor consumes the state to realize a channel & Yes, for $R$ below \\
\bottomrule
\end{tabularx}
\end{table}

\section{Exact state-to-gate compilation by injection}\label{sec:injection}

We now make the ``state as program'' statement operational.  Define the real matrix
\begin{equation}\label{eq:J}
 J=XZ=
 \begin{pmatrix}0&-1\\1&0\end{pmatrix},
 \qquad J^2=-I.
\end{equation}
For $\theta\in\R$, let
\begin{equation}\label{eq:Rtheta}
 R_\theta=e^{\theta J}=\cos\theta\,I+\sin\theta\,J,
 \qquad
 M_\theta:=R_\theta\ket0
 =\cos\theta\ket0+\sin\theta\ket1.
\end{equation}
The target state of the paper is $M=M_{\pi/8}$ and the target gate is
\[
 R:=R_{\pi/8}=e^{\pi J/8}.
\]

Let $C_J$ denote controlled-$J$ with the helper qubit as control and the data register as target:
\[
 C_J=\proj0\otimes I+\proj1\otimes J.
\]
This is a Clifford operation: with the same control and target, $C_J=C_XC_Z$, since the control-$1$ block is $XZ=J$.  The remaining ingredients below---Hadamard, computational-basis measurement, and conditional Clifford correction in the $\theta=\pi/8$ case---are likewise free in the stabilizer model.

\begin{theorem}[Injection identity]\label{thm:injection}
For every data state $\ket\psi$,
\begin{equation}\label{eq:injection}
 (H\otimes I)C_J\left(M_\theta\otimes\ket\psi\right)
 =\frac1{\sqrt2}
 \left(
 \ket0\,R_\theta\ket\psi
 +\ket1\,R_\theta^{-1}\ket\psi
 \right).
\end{equation}
Each measurement outcome on the helper occurs with probability $1/2$.  If $R_\theta^2$ is available as a free correction, one copy of $M_\theta$ implements $R_\theta$ deterministically on arbitrary data.
\end{theorem}

\begin{proof}
From \eqref{eq:Rtheta}, $M_\theta=c\ket0+s\ket1$ with $c=\cos\theta$, $s=\sin\theta$.  Controlled-$J$ gives
\[
 c\ket0\ket\psi+s\ket1J\ket\psi.
\]
Applying Hadamard to the helper yields
\[
 \frac1{\sqrt2}\left[
 \ket0(cI+sJ)\ket\psi+\ket1(cI-sJ)\ket\psi
 \right].
\]
Since $J^2=-I$, $e^{\theta J}=cI+sJ$ and $e^{-\theta J}=cI-sJ$, proving \eqref{eq:injection}.  Both branch operators are unitary and have prefactor $1/\sqrt2$, so the outcomes are equiprobable.  On outcome $1$, applying $R_\theta^2$ gives $R_\theta^2R_\theta^{-1}=R_\theta$.
\end{proof}

At the Hadamard magic angle,
\begin{equation}\label{eq:correction}
 R^2=e^{\pi J/4}
 =\frac1{\sqrt2}(I+J)
 =HZ=XH,
\end{equation}
which is Clifford.  Thus the entire processor outside the consumed state $M$ is Clifford.

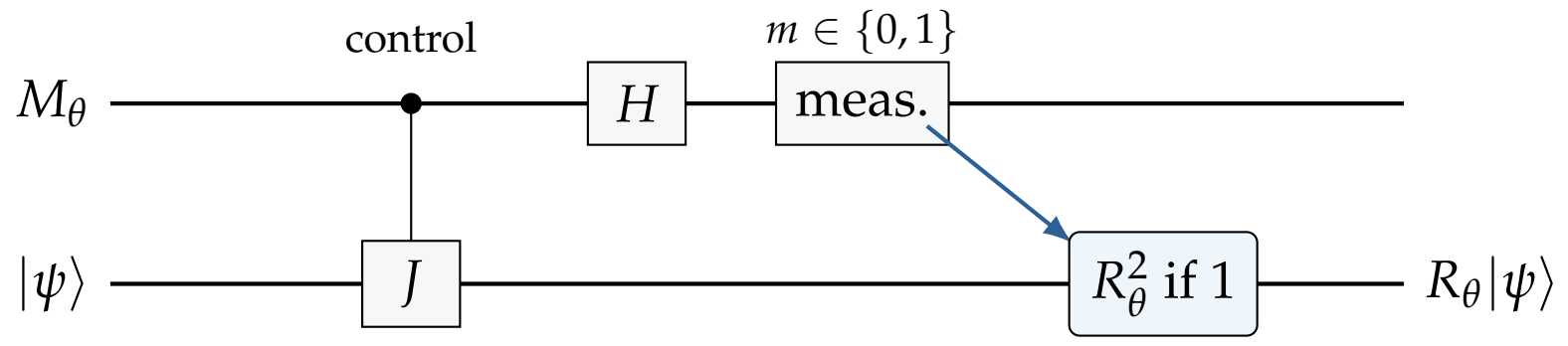
\begin{figure}[t]
\centering
\begin{tikzpicture}[x=1cm,y=1cm, every node/.style={font=\small}]
\draw[thick] (0,1.2)--(8.6,1.2);
\draw[thick] (0,0)--(8.6,0);
\node[left] at (0,1.2) {$M_\theta$};
\node[left] at (0,0) {$\ket\psi$};
\fill (2.0,1.2) circle (2pt);
\draw (2.0,1.2)--(2.0,0);
\node[draw,minimum width=0.65cm,minimum height=0.55cm,fill=softgray] at (2.0,0) {$J$};
\node[draw,minimum width=0.65cm,minimum height=0.55cm,fill=softgray] at (3.5,1.2) {$H$};
\node[draw,minimum width=0.9cm,minimum height=0.55cm,fill=softgray] at (5.0,1.2) {meas.};
\draw[-{Latex[length=2mm]},thick,draw=figblue] (5.43,1.05) -- (6.4,0.27);
\node[draw,rounded corners=2pt,minimum width=1.25cm,minimum height=0.55cm,fill=theoremblue] at (7.0,0) {$R_\theta^2$ if $1$};
\node[right] at (8.6,0) {$R_\theta\ket\psi$};
\node[above,font=\scriptsize] at (2.0,1.4) {control};
\node[above,font=\scriptsize] at (5.0,1.4) {$m\in\{0,1\}$};
\end{tikzpicture}
\caption{State-to-gate injection.  Outcome $0$ applies $R_\theta$ directly; outcome $1$ applies $R_\theta^{-1}$ and is repaired by $R_\theta^2$.  For $\theta=\pi/8$ the correction is Clifford.}
\label{fig:injection}
\end{figure}

\section{One-shot resource duality and universality}\label{sec:duality}

The injection theorem gives one direction of a conversion.  The reverse direction is immediate from the definition:
\begin{equation}\label{eq:prep}
  R_\theta\ket0=M_\theta.
\end{equation}
To avoid a misleading claim that a state and a reusable gate are literally identical resources, we formulate a one-shot relation.

\begin{definition}[One-shot free conversion]\label{def:oneshot}
Fix a class of free operations $\mathfrak F$ containing state preparation of $\ket0$, Clifford unitaries, computational-basis measurement, and classical feed-forward.  We write $A\succeq_{\mathfrak F}^{(1)}B$ when one use or one copy of resource $A$, together with operations in $\mathfrak F$, realizes one use or one copy of $B$.
\end{definition}

\begin{theorem}[One-shot state--gate duality]\label{thm:duality}
Let $M=M_{\pi/8}$ and $R=R_{\pi/8}$.  Relative to Clifford processing,
\begin{equation}\label{eq:duality}
 M\simeq_{\mathrm{Cliff}}^{(1)}R.
\end{equation}
More explicitly, one invocation of $R$ prepares one copy of $M$ from $\ket0$, and one copy of $M$ implements one invocation of $R$ on arbitrary data.
\end{theorem}

\begin{proof}
The conversion $R\to M$ is \eqref{eq:prep}.  The conversion $M\to R$ is \cref{thm:injection} together with the Clifford correction \eqref{eq:correction}.
\end{proof}

This is the precise sense in which the state and gate are interchangeable resources.  It is an operational equivalence in a declared processor model, not an identification of Hilbert-space vectors with operators.

\subsection{Universality consequence}
The gate $R$ is a real single-qubit rotation that does not preserve the computational basis and is non-Clifford.  Shi proved that CNOT together with any suitable real one-qubit gate of this kind yields universal quantum computation over real amplitudes, apart from the explicitly characterized exceptional cases \cite{Shi2003}.  Consequently a real Clifford processor supplied with copies of $M$ can inject $R$ and thereby reach universal real quantum computation in the usual approximation sense.

There is also a direct bridge to the conventional Clifford+$T$ picture.  Put $C=SH$.  Since $C Z C^\dagger=Y$ and $T=e^{i\pi/8}e^{-i\pi Z/8}$,
\begin{equation}\label{eq:RcliffordT}
 R=e^{-i\pi Y/8}=e^{-i\pi/8}\,C T C^\dagger.
\end{equation}
Thus access to $R$ together with the full Clifford group is equivalent, up to global phase and Clifford conjugation, to access to $T$.  The familiar Clifford+$T$ universality therefore appears in the same processor model.

\begin{corollary}[Magic misty resource]\label{cor:universality}
A supply of the nested misty state $M=\Norm\{W,\Norm\{W,B\}\}$ promotes the declared Clifford injection processor to a universal quantum resource: it gives universal real computation using the real gate fragment, and, with the full Clifford group, recovers Clifford+$T$ universality through \eqref{eq:RcliffordT}.
\end{corollary}

The corollary does not assert that the misty notation is a new route to the established universality of magic-state computation.  Its point is representational: a nested source-language value discovered as a fixed point is simultaneously a consumable program resource.

\section{A programming-language semantics for dual-role values}\label{sec:pl}

The two roles of $M$ can be stated as semantic certificates rather than informal labels.  Let $\llbracket t\rrbracket$ denote the state semantics of a closed misty term $t$.

\begin{definition}[Invariance certificate]\label{def:fixcert}
A judgment
\[
 \vdash t:\Fix(U)
\]
is sound when
\[
 U\llbracket t\rrbracket=\llbracket t\rrbracket.
\]
\end{definition}

\begin{definition}[Program-resource certificate]\label{def:progcert}
A judgment
\[
 \vdash t:\Prog(R;\cI)
\]
is sound when the declared injection processor $\cI$, after consuming one copy of $\llbracket t\rrbracket$ and applying its allowed feed-forward, realizes the unitary channel
\[
 \cU_R(\rho)=R\rho R^\dagger
\]
on arbitrary input states, including inputs entangled with an external reference.
\end{definition}

Because \cref{thm:injection} proves equality at the operator level branch by branch, the program-resource certificate is stable under extension by an untouched reference system.

\begin{theorem}[Dual-role soundness]\label{thm:dualrole}
For
\[
 t_M=\Norm\{W,\Norm\{W,B\}\},
\]
we have both
\[
 \vdash t_M:\Fix(H)
 \qquad\text{and}\qquad
 \vdash t_M:\Prog(R;\cI_{CJ}),
\]
where $\cI_{CJ}$ is the controlled-$J$ processor of \cref{fig:injection} with Clifford feed-forward.
\end{theorem}

\begin{proof}
The first judgment is \cref{cor:involution} with $U=H$ and $x=\ket0$.  The second is \cref{thm:injection,eq:correction}.
\end{proof}

The judgments classify different semantic facts.  $\Fix(H)$ is extensional state invariance; $\Prog(R)$ is an operational statement about what an external processor does when the state is supplied and consumed.  A richer language could track both as refinement/effect information without collapsing them into one type.

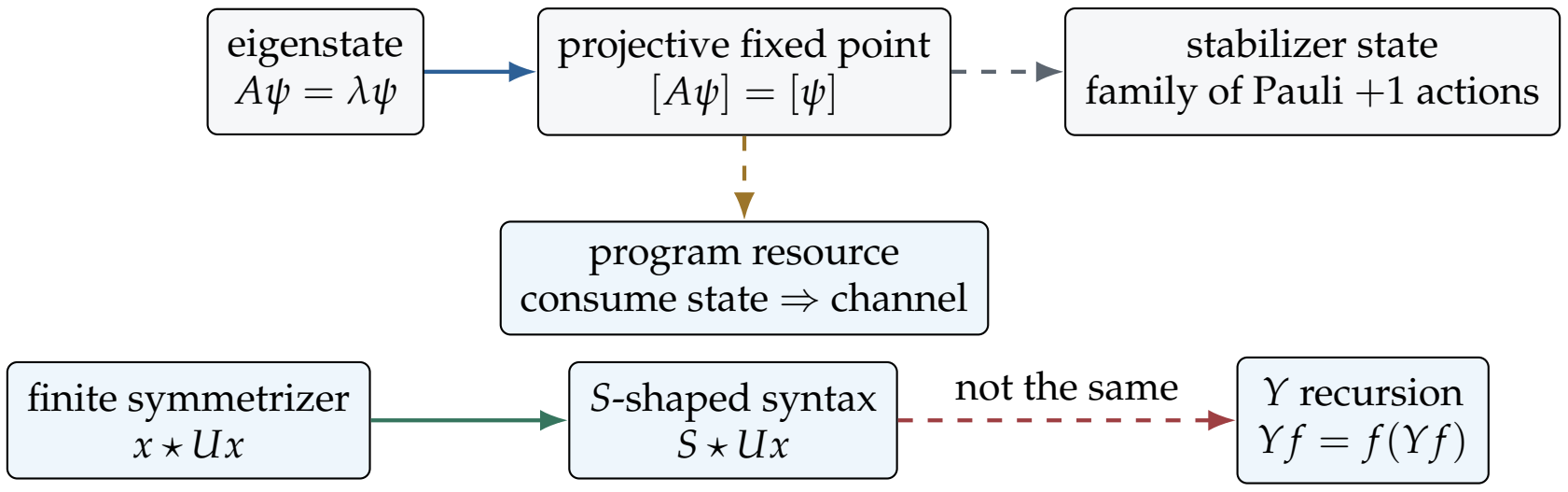
\begin{figure}[t]
\centering
\begin{tikzpicture}[node distance=6mm and 8mm, every node/.style={font=\scriptsize},
 box/.style={draw=black,rounded corners=2pt,align=center,inner sep=4pt,minimum height=8mm},
 arr/.style={-{Latex[length=2mm]},thick}]
\node[box,fill=softgray] (eig) {eigenstate\\$A\psi=\lambda\psi$};
\node[box,right=of eig,fill=softgray] (proj) {projective fixed point\\$[A\psi]=[\psi]$};
\node[box,right=of proj,fill=softgray] (stab) {stabilizer state\\family of Pauli $+1$ actions};
\node[box,below=of eig,fill=theoremblue, yshift=-10mm, xshift=-9mm] (sym) {finite symmetrizer\\$x\star Ux$};
\node[box,right=of sym,fill=theoremblue, xshift=6mm] (s) {$S$-shaped syntax\\$S\star Ux$};
\node[box,right=of s,fill=theoremblue, xshift=16mm] (y) {$Y$ recursion\\$Yf=f(Yf)$};
\node[box,below=of proj,fill=theoremblue] (prog) {program resource\\consume state $\Rightarrow$ channel};
\draw[arr,draw=figblue] (eig)--(proj);
\draw[arr,draw=figgray,dashed] (proj)--(stab);
\draw[arr,draw=figgreen] (sym)--(s);
\draw[arr,draw=figred,dashed] (s)-- node[above]{not the same} (y);
\draw[arr,draw=figgold, dashed] (proj)--(prog);
\end{tikzpicture}
\caption{The conceptual neighborhood.  Solid arrows mark precise implications or constructions used here; dashed links indicate comparisons that require additional hypotheses and must not be read as equivalences.}
\label{fig:taxonomy}
\end{figure}

\section{Combinators, recursion, and the boundary of the analogy}\label{sec:combinators}

The fixed-point constructor has an exact syntactic relation to the $S$ combinator.  In curried combinatory logic,
\begin{equation}\label{eq:S}
 Sfgx=fx(gx).
\end{equation}
If a binary constructor $\star$ is treated as $f$ and a unary operation $U$ as $g$, then
\begin{equation}\label{eq:Sstar}
 S\star Ux=\star x(Ux)=x\star Ux.
\end{equation}
Thus the misty constructor has an $S$-shaped \emph{sharing pattern}: the same argument $x$ is supplied both directly and through $U$.

The semantic fixed-point property, however, does not come from the defining equation of $S$.  It comes from the algebraic hypotheses $U^2=I$, symmetry of $\star$, and equivariance \eqref{eq:equivariance}.  This is precisely why $S$ and $Y$ should not be conflated.

A fixed-point combinator $Y$ satisfies
\begin{equation}\label{eq:Y}
 Yf=f(Yf),
\end{equation}
so $Y$ is a general mechanism for obtaining a recursive fixed point of $f$ in a calculus where such self-application is meaningful \cite{Barendregt1984}.  Denotational semantics often interprets recursive definitions through least fixed points of continuous operators on ordered domains \cite{Winskel1993}.  Neither construction is the same as finite group averaging.

The distinction can be summarized as follows.
\begin{itemize}
\item $x\star Ux$ is a finite algebraic symmetrizer.  It constructs a fixed state because a finite orbit is averaged.
\item $S\star Ux$ exposes the sharing/data-flow pattern of that expression.
\item $Yf$ is a recursive fixed-point constructor whose defining equation is itself self-referential.
\item An attractor is a dynamical notion: nearby states converge under iteration or continuous-time evolution.  A unitary eigenstate need not be attracting; in particular $H^2=I$ causes generic states to oscillate rather than converge.
\end{itemize}

This separation is not cosmetic.  It prevents the common but incorrect inference that a quantum eigenstate, a stabilizer state, a recursive fixed point, and an attractor are merely different names for the same mathematical object.  What they share is an invariance schema; the mechanisms that establish or exploit the invariance are different.

\section{Scope, prior work, and claim boundaries}\label{sec:scope}

The results sit at the intersection of several mature subjects, so the claim boundary is part of the theorem architecture.

\paragraph{Misty-state semantics.}
Rudolph introduced the black-and-white misty-state representation as an accessible rewriting language for quantum computation \cite{Rudolph2017}.  German's 2026 work explicitly developed irreducible/fixed-point misty states and used them in a term-rewriting semantics \cite{German2026}.  Newton and German supplied the explicit normalization discipline used here \cite{NewtonGerman2026}.  We do not claim to introduce the misty notation, the nested Hadamard state, or the basic idea of fixed points in that notation.

\paragraph{Magic states and gate injection.}
Magic-state computation and the sufficiency of nonstabilizer ancillary states with Clifford processing are established by Bravyi and Kitaev and related gate-teleportation work \cite{BravyiKitaev2005,GottesmanChuang1999}.  We do not claim the discovery of magic-state injection or of H-type magic states.  The contribution is the exact processor theorem \cref{thm:injection} stated in the real misty $J=XZ$ representation and its use to formalize a dual-role misty term.

\paragraph{Universality.}
The universality statement relies on established real-gate universality results, in particular Shi's characterization of real single-qubit gates that promote classical/restricted gate sets to universality \cite{Shi2003}.  Our corollary simply transfers that result through the exact state-to-gate injection.

\paragraph{Fixed points and combinators.}
The equations for $S$ and $Y$, and the use of fixed points in denotational semantics, are classical programming-language theory \cite{Barendregt1984,Winskel1993}.  The paper does not identify those constructions with quantum eigenstates.  Instead, it isolates one exact syntactic sharing identity \eqref{eq:Sstar} and one exact semantic contrast: finite orbit symmetrization versus recursive fixed-point construction.

\paragraph{What is new in the present synthesis.}
Within the declared misty-state fragment, the main contribution is a single auditable chain:
\[
\boxed{\begin{gathered}
\text{normalized mist syntax}\longrightarrow\text{certified finite-orbit fixed point}\\
\longrightarrow\text{H-type magic state}\longrightarrow\text{one-shot gate program}
\end{gathered}}
\]
The fixed-point and program roles are carried by the same denotation but certified by different semantic judgments.  This makes the ``programs are data'' analogy precise at one nontrivial quantum resource interface without asserting a universal state--gate correspondence.

\subsection{Conclusion and next theorem frontier}\label{sec:conclusion}

The nested misty term
\[
 M=\Norm\{W,\Norm\{W,B\}\}
\]
is a compact example of how one quantum object can legitimately inhabit several mathematical roles without those roles becoming synonymous.  As a state it is the $+1$ Hadamard eigenvector.  As the two-element orbit cloud of $\ket0$ under $H$, it is a certified fixed point generated by finite symmetrization.  As a nonstabilizer H-type state, it is a magic resource.  As an input to the controlled-$J$ processor, it is consumed to implement the non-Clifford gate $R=e^{\pi XZ/8}$ on arbitrary data.  Conversely, one invocation of $R$ prepares the state.  The exact operational statement is therefore the one-shot resource duality
\[
 \boxed{M\simeq_{\mathrm{Cliff}}^{(1)}R.}
\]

The programming-language lesson is equally specific.  The term can carry both an invariance certificate $\Fix(H)$ and a program-resource certificate $\Prog(R)$; these are distinct judgments about the same denotation.  The constructor $x\star Ux$ has an exact $S$-combinator sharing pattern, but its fixed-point property comes from finite symmetry, not from recursion.  $Y$ remains the genuinely recursive mechanism.  This separation turns a suggestive analogy into a small semantic discipline.

The next theorem frontier is to generalize the program-resource certificate from the single H-type resource to a typed family of dyadic mist terms, where the syntax records both the target gate and the adaptive correction structure.  Such an extension should preserve the central discipline of the present paper: every claimed program interpretation must be backed by an exact channel identity, and every invariance claim by an explicit semantic certificate.

\section*{Acknowledgments} 
We thank Prof. Terry Rudolph for helpful conversations about this work.

\bibliographystyle{plain}

\vfill

\end{document}